**Title**: Impact of Environmental Colors on Human Aggressiveness: Insights from a Minecraft-Based Behavioral Study

**Authorship**: Austin Deng-Yao Yang[1], Shih-Jen Tsai[2,3], Hsin-Jung Tsai[3,*]

**Affiliation**:

[1] International Bilingual High School in Hsinchu Science Park, Hsinchu, Taiwan

[2] Department of Psychiatry, Taipei Veterans General Hospital, Taipei, Taiwan

[3] Institute of Brain Science, National Yang-Ming Chiao Tung University, Taipei, Taiwan

*Correspondence to **Hsin-Jung Tsai**, Institute of Brain Science, National Yang-Ming Chiao Tung University, No. 155, Sec. 2, Linong St., Beitou Dist., Taipei 11221, Taiwan. Tel: +886-2-2826-7054; E-mail: hjtsai2@nycu.edu.tw

**Running Title:** Color and Behavior

**Word count**: Abstract (256 words); Body (3,266 words); 2 Tables; 4 Figures

**Conflict of Interests**:  None of the authors has any potential conflict of interest to be disclosed.

**Acknowledgments:** We thank all the anonymous players on Minecraft who participated in this study.

**Funding Information:** This study was not funded.

**Abstract**

This study explores the influence of environmental colors on human behavior, specifically focusing on aggressiveness and passiveness. Color is widely regarded as an influential environmental factor shaping human behavior, yet existing studies present conflicting evidence regarding its impact on aggressiveness and passiveness. This study employed Minecraft as a controlled digital platform to investigate whether exposure to different colors influences both the frequency and nature of participant interactions (aggressive versus non-aggressive), and whether prolonged exposure amplifies these effects. Anonymous online participants were exposed to various colors before interacting with non-player characters simulating human-like encounters. Three key outcomes were measured: (1) total interactions per color, (2) ratios of aggressive to non-aggressive interactions per color, and (3) the effect of varying exposure durations on aggressiveness. While no significant overall differences in interaction frequency were observed among the colors, post-hoc analyses revealed that Red and Black elicited significantly more interactions compared to Green. Additionally, Red, Yellow, and Black were associated with higher ratios of aggressive behavior relative to Green or White. Prolonged exposure to Red also appeared to intensify aggressive responses. These findings underscore the potential role of environmental color in shaping online social behaviors and highlight the importance of environmental settings in areas ranging from online communication platforms to digital marketing strategies.

1. **Introduction**

The environment is a complicated and prominent influence that affects humans in many ways. Human behavior is very complex that encompasses a vast plethora of different things. From the evolutionary aspects (Jalil et al., 2012) to personal experience (Piga et al., 2021), studies have also reported that environmental inputs of color affect human emotion and behavior (AL-Ayash et al., 2016). Exposure to different colors has been linked to different types of reactions (Jalil et al., 2012). By using the color of the environment as a variable, the experiment can effectively simulate changes in the environment and induce a reaction in participants.

A field experiment exploring whether the color of cars was related to an aggressive response was conducted, with test drivers waiting at a traffic light, where it was blocked by an experimental car of varying colors. This study found that the red car elicited an earlier aggressive response (Guéguen et al., 2012). However, inconsistent results were reported in another study using online settings (Guéguen et al., 2012). The effect of the color red on risk-taking behavior was tested in two web-based experiments. One of these experiments had results with a behavioral outcome: Respondents showed more cautious and less risk-taking behavior in a game when the main focal stimuli were red objects, which showed that red color leads to more conservative choices and behaviors, contrary to the expected aggressive behavior commonly linked to the color red, compared with blue objects (Gnambs et al., 2015).

In the digital age, anonymous online platforms provide unique circumstances where social behavior could significantly differ from in-person interaction. Online social behavior places the act of socializing in a radically different environment, and thus behavior in such a medium would receive a myriad of different factors otherwise unknown in the physical world. Likewise, many factors that influence social behavior in-person are often nonexistent in the context of online social behavior. It is often the case in which an individual acts much differently over text messages versus communicating in-person (Spagnolli and Gamberini, 2007; Holtgraves and Paul, 2013).

      Whether changes in the color of the environmental elements of the online platform as an input will induce a chain of reactions, including the number of online social interactions, and the number of interactions involved in aggressiveness or passiveness, remains uncertain. Moreover, whether it amplifies aggressiveness by the duration of exposure to specific colors in digital platforms is an important variable to determine the role of color exposure on internet-based social behavior. Because it directly reflects a commonly found aspect of the effect of environments on humans, which is the fact that when a person gets exposed to a certain environment for a longer time, they are more likely to be differently or more drastically affected by the environment than a person who is exposed to the environment for a shorter length in time.

      Thus, the rise of online platforms introduces additional layers of complexity in studying social and behavioral dynamics. Individuals often behave differently in virtual or anonymous digital contexts than in face-to-face interactions. This shift calls for further exploration into whether environmental factors—specifically, color—still exert comparable effects in online settings, and whether the duration of exposure to certain colors amplifies these effects. Minecraft (https://www.minecraft.net/) offers a versatile and controlled digital environment for investigating these questions. Originally designed as a single-player, open-world survival game, Minecraft has evolved into a platform that

supports multi-user "worlds," thus enabling real-time observation of group behaviors. The current study leverages this environment to examine how different background colors affect online social interactions, including the balance between aggressive and non-aggressive responses. We hypothesized that more vivid or deeper colors (e.g., red or black) will elicit higher levels of hostile behavior than lighter, more neutral colors (e.g., green or white), and that these tendencies would be further magnified by prolonged exposure.

## 2. Materials and Methods

The current study was conducted around May to the start of June in 2021, with a group of anonymous, real players throughout the Minecraft community of different servers worldwide (e.g., Hypixel) joining this study.

*2.1 Experimental Setting*

This study comprised experiments for changing colors or durations of exposure to identify the role of color exposure and social behavior (i.e., the number of interactions and the associated content) through an online platform. For the color exposure experiment, participants were exposed to the environment under different color settings that were generated in a random order (i.e., color HEX codes: red as #FF0000, yellow as #FFFF00, blue as #0000FF, pink as #FF00FF, black as #000000, or white as #FFFFFF). These colors may vary slightly when presented to each participant, depending on the device used. In this experiment where the only changing variable was the color, each participant was exposed to their given color for around 5 seconds.

Moreover, to determine the effect of color exposure duration on online social behaviors, the default color was set to red, and instead of changing the color. Participants were then exposed to the red color for different amounts of time for 2 s, 5 s, 10 s, or 30 s per test. While the test subjects were not restricted from leaving the starting area before the specified time, in all cases the participant had followed the given directions to wait their respective amounts of seconds before proceeding.

The layout of the experiment was set as follows: each of the participants was exposed to a certain color for a set amount of time in a confined room, then they walked along a designated pathway to another room. At the start, participants received two items, a blue sword meant to be used as a weapon, and an emerald, that are part of the test. During the process, they were allowed to interact using the given items, and the items represented aggressiveness and friendliness, respectively, with Minecraft entities known as villagers set at predetermined places along the hallway, for a total of 3 villagers. An interaction using the sword would involve the participant utilizing a mouse click while holding the item, resulting in the villager being killed visually by the in-game player with the sword. An interaction with the emerald would involve the player dropping the item in the vicinity of one of the villagers. These villagers are non-player-characters that coded into the experiment to simulate real players. The participants would not be told whether or not to interact with the villagers, they only received the instruction to stay at the room on the other end of the pathway after they are done passing through the hallway. Figure 1 illustrated a visual representation of the actual layout of the Minecraft environment used in this study.

*2.2 Data Collection*

To determine whether color exposure is associated with online social behavior, the number and the associated content for each participant under the experiments were recorded.

The number of interactions that participants had with the villagers was first collected. There were 3 villagers per test, and the number of interactions for each participant was set from 0 to 3. An interaction was recorded when the participants once approached the villager, including using one of the items given to them from the chest at the start.

On the other hand, the content of interactions under different color exposure that the participants had with the villagers was also recorded. Each interaction was characterized as aggressiveness (e.g., when the subject attacks any of the villagers) or non-aggressive behavior (i.e., when the subject gives an emerald to the villager or takes no specific action at all). This method is used as it allows for the gathering of data corresponding to which colors result in aggressive behavior and which colors result in passive behavior. White, green, and red are used as control groups; white is commonly regarded as a neutral color with no saturation, green is the "least aggressive" color, and red is the "most aggressive" color.

*2.3 Statistical analysis*

The statistical analyses followed the three parts involved in this study. First, the analysis of variance (ANOVA) was employed to assess the effect of color exposure on the number of interactions between participants and villagers. Then, the number of interactions under each color (i.e., red, yellow, blue, pink, black, or white) paired with the control color of green or red formed the color pairs. Post-hoc comparisons between pairs of colors were conducted to analyze the differences in the number of interactions pertaining to the different color pairs. Second, the chi-square test was used to assess the difference in the ratio of the type of behavior with regards to the different color pairs. Third, a trend test was used to evaluate the significance of the trend of the number of aggressive behaviors with regards to the duration of exposure to the red color. A *p*-value of less than 0.05 was considered statistically significant.

## 3. Results

*3.1 Participant*

Since the experiments were conducted on open software Minecraft, it is impractical to verify the identities, ages, and gender of all the participants who participated. Hence, participants' demographics were not collected.

A group of 15 participants was enrolled in the first experiment to quantify the total number of interactions. For the second experiment, 20 participants were recruited to evaluate the presence of aggressive behavior. In total, 35 anonymous individuals participated in the study. Of these, 10 participants volunteered for the following test to further investigated the impact of exposure duration on behavioral patterns.

*3.2 The effect of color on the interactions between the test subject and environment*

In the first experiment, participants were initially exposed to different colors and the number of interactions with the villagers during the test followed by the exposure was counted (Figure 2). The between-group analysis, conducted using ANOVA, showed no significant differences in the number of interactions among various colors (F = 1.46, $P$ = 0.20). Tables 1 and 2 present post-hoc analyses of paired group comparisons, with Green and Red serving as control groups, respectively. These results revealed a significantly higher number of interactions under the colors Red ($t$ = 2.331, $P$ = 0.015) and Black ($t$ = 2.087, $P$ = 0.049), compared to Green. Furthermore, comparisons between other colors and Red demonstrated significantly fewer interactions for Green ($t$ = -2.331, $P$ = 0.015), Pink ($t$ = -1.750, $P$ = 0.048), and White ($t$ = -1.826, $P$ = 0.041).

*3.3 The effect of color on the type of interaction*

As the total number of interactions per color are different, a ratio between the number of aggressive interactions and the number of non-aggressive interactions under different color settings were calculated (Figure 3). The main effect was found by a chi-square statistic to compare the ratio of aggressive and non-aggressive interactions under seven color settings ($\chi^2 = 23.37$, $P = 0.0007$). Within that, color of Green (aggressiveness to non-aggressiveness, 3/16 = 0.19) or White (aggressiveness to non-aggressiveness, 5/16 = 0.31) had the least ratio of aggressiveness to non-aggressiveness, and both colors were then being used to compare with others to perform post-hoc analysis. The color White is regarded as a control group, and will likewise be compared relatively to other colors, primarily because it is widely considered a "neutral color" with a saturation value of 0.

The results suggested that Red, Yellow, and Black had significantly higher ratios of aggressiveness to non-aggressiveness, compared to Green or White, while Pink and Blue show no significant difference in the ratios of aggressiveness to non-aggressiveness. Specifically, comparing red to green brings us a Chi-square of 8.2 ($P = 0.004$). Similarly, Red to White results in a Chi-square of 5.72 ($P = 0.017$). Comparing Yellow gives similar results; Yellow to Green has a Chi-square of 9.86 ($P = 0.002$) while Yellow to White has a chi-square of 7.25 ($P = 0.007$). A comparison for Black can also be made. Black to Green gives a Chi-square of 10.32 ($P = 0.001$) and Black to White gives a Chi-square of 7.64 ($P = 0.006$).

*3.4 The effect of duration of exposure on the aggressiveness*

This test was conducted using a separate group of people from the previous two tests. The participants were split into four groups with an equal amount of people, with each group representing a set exposure time (2 seconds, 5 seconds, and so on) The test was then conducted as usual, and the amount of aggressive behavior was recorded, and the total amount of aggressive behavior was noted (Figure 4). Visual inspection indicated a trend of escalating aggressiveness correlated with prolonged exposure duration. However, there were not enough participants to determine a correlation.

**4. Discussion**

This study yielded three key results: (1) While overall interactions did not vary significantly across colors, post-hoc analysis revealed that the colors Red and Black elicited significantly more interactions compared to Green; (2) Higher ratios of aggressive versus non-aggressive interactions were found under the colors Red, Yellow, and Black, particularly when compared to the less aggressive responses associated with Green and White; and (3) Preliminary observations suggested a trend where longer exposure to colors correlated with increased aggressiveness. These findings suggest that color exposure may influence human behavior, particularly in terms of interaction type and intensity. The distinct behavioral responses to different colors, especially the heightened interactions and aggressiveness under certain colors, highlight the potential of environmental color as a factor in influencing human behavior. This has implications for environments where color is a controllable element, such as in therapeutic settings, educational environments, or marketing.

*4.1 Impact of color on interactions between participants and environment*

The study's findings regarding the impact of color on the interactions between participants and their environment offer intriguing insights. The initial exposure to different colors followed by the measurement of interactions with villagers, as depicted in Figure 2, revealed a multifaceted relationship between color and social behavior. Notably, the post-hoc analyses (Tables 1 and 2) highlighted that significantly higher numbers of interactions were noted under Red and Black compared to Green. This suggests that certain colors, particularly those with stronger visual impact like Red and Black, can potentially stimulate more social interactions.

These findings align with previous research indicating that colors can evoke specific psychological responses (Benbasat and Dexter, 1986; Jalil et al., 2012). For example, Red, often associated with alertness and urgency, may have prompted a heightened level of engagement among participants. In contrast, Green, typically linked to calmness and tranquility, appeared to result in fewer interactions. This variation in interaction based on color exposure highlights the potential influence of environmental colors on social behavior.

One previous study demonstrates the role of colors in design, particularly in consumer behavior (Arabi, 2017). This study highlights the power of colors to differentiate products, influence mood, and shape attitudes. Another study provides a cross-cultural perspective, revealing how colors, especially red and black, are imbued with affective meanings that can intensify visual impact and stimulate social interactions (Adams and Osgood, 1973). One more recent research in healthcare settings further supports this notion by demonstrating that color can not only alleviate stress but also enhance patient satisfaction and boost staff morale and productivity (Ghamari and Amor, 2016). Collectively, these studies reinforce the concept that certain colors, especially those with a more pronounced visual impact, are key influencers of social behavior and interaction dynamics.

*4.2 Association of color with type of interaction*

The study further explored the nature of these interactions by examining the ratio of aggressive to non-aggressive interactions under different color settings. The chi-square analysis revealed significant variations in this ratio based on color. Notably, Red, Yellow, and Black were associated with higher ratios of aggressiveness compared to Green or White. In contrast, Pink and Blue showed no significant difference in the ratios of aggressiveness. This observation is particularly interesting as it suggests that not all colors elicit the same type of social response. The higher aggressiveness ratio under Red, Yellow, and Black could be attributed to these colors' association with danger or warning signals in various cultures. This finding is crucial for understanding how color can influence not just the quantity but also the quality of social interactions in digital environments.

Prior research provides mixed findings on the association between environmental color and aggressive behavior. One earlier study addressed various environmental factors that might influence aggression, including color, but the study did not specifically concentrate on the relationship between color and aggression (Anderson, 1982). One recent study focusing on animal behavior specifically examined the effects of light intensity on aggressiveness in a fish species (Sarmento et al., 2017). This research concluded that higher light intensities tend to increase aggression levels. Another research on animal behavior investigates the role of red coloration in aggression in Gouldian finches and suggests that red color may be an innate signal of aggression (Pryke, 2009).

In human study, a prior review focused on stressful environments and aggressive behavior in adolescents, but did not directly address the association between environmental color and aggression (Hudley and Novac, 2007). One earlier study suggested that the color pink can have a calming effect on individuals experiencing anger or agitation (Schauss, 1979). Another study found that black uniforms in professional ice hockey were associated with more aggressive penalties, while white uniforms were associated with fewer penalties (Webster et al., 2012). Additionally, a study found that increased redness in men's faces was associated with perceived aggression and dominance, but not attractiveness (Stephen et al., 2012). This divergence in findings underscores the complexity of establishing a clear-cut relationship between color in the environment and aggressive behavior across different contexts and species.

*4.3 Duration of exposure and aggressiveness*

    The third aspect of the study addressed the effect of the duration of exposure to color on aggressiveness. Although a trend of escalating aggressiveness with prolonged exposure was noted, the lack of sufficient participants in this segment of the study limited the robustness of these findings. However, this observed trend aligns with the notion that prolonged exposure to certain stimuli can amplify the impact of psychosocial behavior. In this context, extended exposure to colors attributive to aggression might exacerbate such behavioral tendencies.

    Overall, this experiment is an example of machine reductionism. The definition of machine reductionism is a method of explaining behavior as an analogy with simpler machine systems. In this example, human aggressiveness and behavior in general is extremely complex, but in the experiment, the various choices and outcomes are simplified, but the experiment still simulates and encompasses most aspects of human behavior. As is, machine reductionism was used in this experiment to answer the original hypothesis and provide more insight on human behavior, as well as improving the efficiency of the test.

*4.4 Limitations*

Despite these insightful findings, this pilot study is subject to several limitations. First, because the study relied on anonymous volunteers within the Minecraft community, it was impossible to verify participants' demographic information such as age, gender, or cultural background. This limitation restricts the generalizability of the findings across broader populations, as different demographic groups may respond differently to color cues. Second, the study featured a relatively small sample size, especially for the subgroup examining prolonged color exposure. This limited number of participants reduced the statistical power to detect strong relationships or draw definitive conclusions regarding how extended exposure to specific colors influences aggression.

Third, the unique digital setting of Minecraft, although advantageous for controlled experimentation, may not fully mirror real-world social dynamics. Participants might behave differently in an online game-based context compared to face-to-face interactions, thus limiting the ecological validity of the results and their applicability to offline environments. Lastly, the digital nature of the experiment introduced device-dependent variations in color representation. Participants viewed colors on various devices—monitors, laptops, or tablets—with potentially divergent display settings, which may have slightly altered the intended color exposure conditions.

Future research should address these limitations by employing larger, more diverse samples, standardized color calibration methods, and alternative platforms or real-world contexts to enhance the robustness and applicability of the findings.

*4.5 Conclusions*

The current study employed a real-world design in which all the participants were anonymous users on one of the world's largest online platforms. Our findings provide initial evidence that color exposure can significantly influence social behaviors in a digital setting. Specifically, Red and Black were associated with increased interaction frequencies, while Red, Yellow, and Black elicited more aggressive behaviors compared to neutral or calming colors, such as Green and White. These findings underscore the potential of color as a modifiable factor in shaping user engagement and affective responses in virtual environments. Future research with larger and more diverse participant samples, as well as extended exposure durations, is warranted to confirm and expand upon these results. Our findings further offer valuable insights into the psychological impact of color in digital settings, suggesting potential applications in areas ranging from online communication platforms to digital marketing strategies.

**Figure Legends**

**Figure 1.** A demonstration of experimental layout. One participant walked through the designated pathway to the other room, receiving items from the chest and interacting with villagers.

**Figure 2.** Number of interactions showed between participants and villagers during experiments for each color from all the participants. Data were expressed as mean and standard deviation (error bar).

**Figure 3.** Ratios of aggressiveness to non-aggressiveness for each color.

**Figure 4.** Trends in total aggressive interactions associated with the color red across various exposure times.

**Table 1** Two-Sample t-Test Analyses Comparing Number of Interactions Under Green Color Sample With Each Other Color

| Color Pair | *t* | *P* value |
| --- | --- | --- |
| Red / Green | 2.331 | 0.015 |
| Yellow / Green | 1.268 | 0.219 |
| Blue / Green | 1.335 | 0.197 |
| Pink / Green | 0.607 | 0.550 |
| Black / Green | 2.087 | 0.049 |
| White / Green | 0.381 | 0.707 |

**Table 2** Two-Sample t-Test Analyses Comparing Number of Interactions Under Red Color Sample With Each Other Color

| Color Pair | t | P value |
|---|---|---|
| Yellow / Red | -1.069 | 0.14 |
| Green / Red | -2.331 | 0.015 |
| Blue / Red | -1.164 | 0.129 |
| Pink / Red | -1.750 | 0.048 |
| Black / Red | -0.313 | 0.379 |
| White / Red | -1.826 | 0.041 |

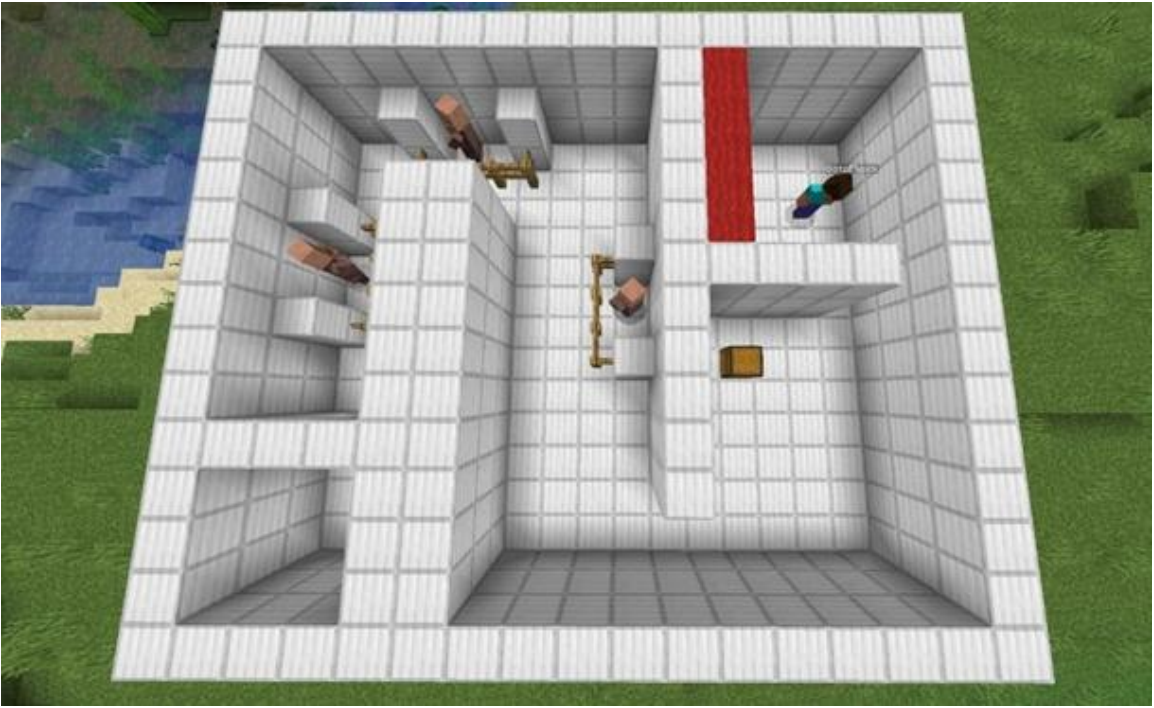

**Figure 1.**

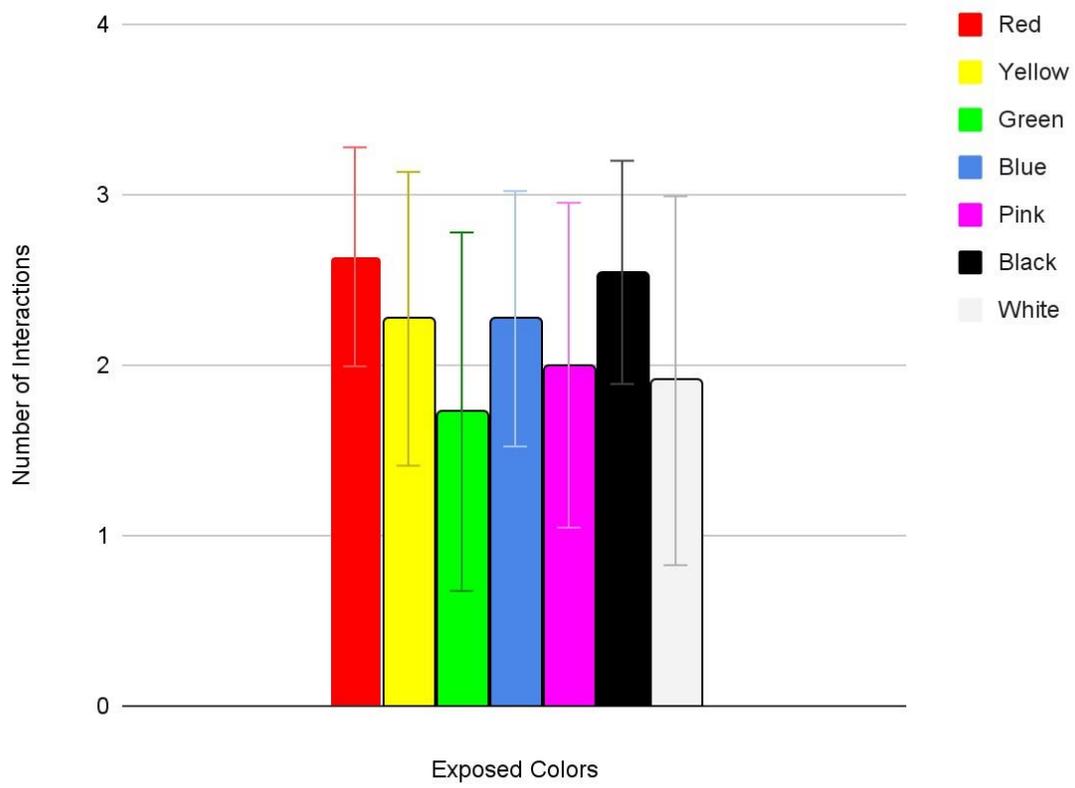

**Figure 2.**

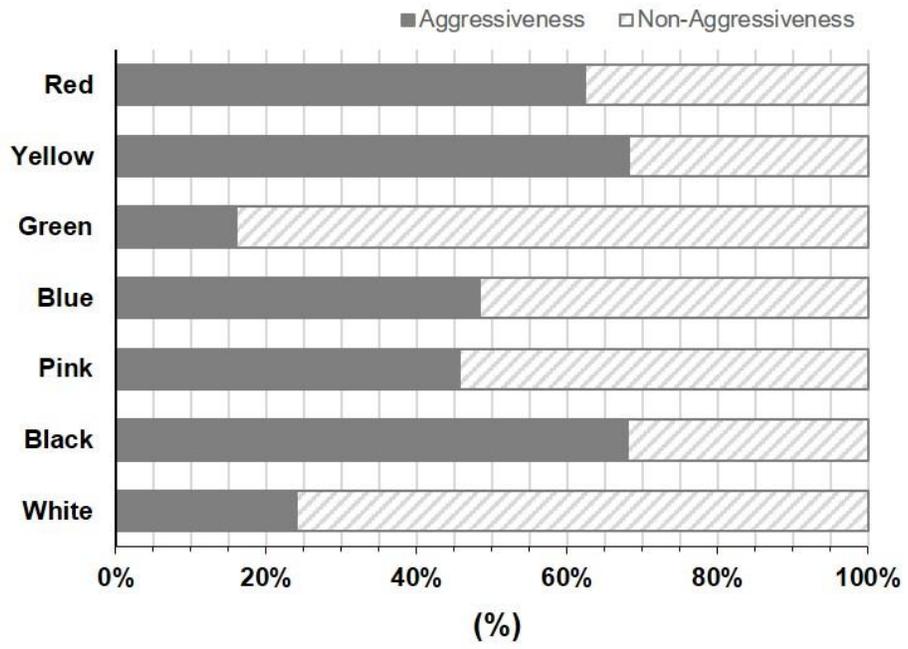

**Figure 3.**

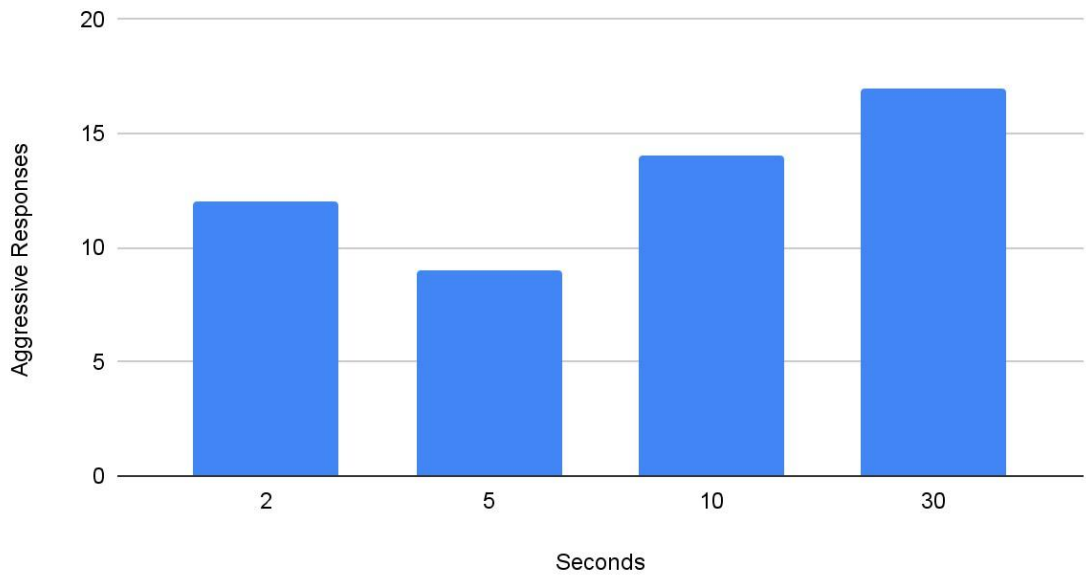

**Figure 4.**